\documentclass[11pt]{article}
\usepackage{epsfig}
\usepackage{aaspp4}
\usepackage{amstex}

\begin{document}
\title{Magnetized Iron Atmospheres for Neutron Stars}
\author{Mohan Rajagopal\altaffilmark{1}, Roger W. Romani\altaffilmark{2}}
\affil{Dept. of Physics, Stanford University, Stanford CA  94305-4060}
\authoraddr{Dept. of Physics, Stanford University, Stanford CA  94305-4060}

\author{and M. Coleman Miller\altaffilmark{3}}
\affil{Dept. of Ast. and Astrophys., University of Chicago, Chicago, IL  60637}
\authoraddr{Dept. of Ast. and Astrophys., University of Chicago, 
Chicago, IL  60637 }

\altaffiltext{1}{mohan@@astro.stanford.edu} 
\altaffiltext{2}{Alfred P. Sloan Fellow; rwr@@astro.stanford.edu} 
\altaffiltext{3}{Compton GRO Fellow; miller@@gamma.uchicago.edu} 

\begin{abstract}
Using a Hartree-Fock formalism, we estimate energy levels and photon
cross sections for atomic iron in magnetic fields $B \sim 10^{13}$G.
Computing ionization equilibrium and normal mode opacities with these
data, we construct LTE neutron star model atmospheres at $5.5 < {\rm
Log}(T_{\rm eff})< 6.5$ and compute emergent spectra. We examine the
dependence of the emergent spectra on $T_{\rm eff}$ and $B$. We also
show the spectral variation with the angle between the magnetic field
and the atmosphere normal and describe the significant limb darkening
in the X-ray band. These results are compared with recent detailed
computations of neutron star H model atmospheres in high fields and
with low field Fe and H model atmospheres constructed from detailed
opacities. The large spectral differences for different surface
compositions may be discernible with present X-ray data; we also note
improvements needed to allow comparison of Fe models with high quality
spectra.
\end{abstract}

\section{Introduction}

Neutron star surface emission has long been sought as a measure of the
star's thermal history and, hence, a probe of the equation of state of
matter at supra-nuclear densities. Since temperatures remain near
$10^6$K for $\sim 10^6$y, this emission is best probed in the soft
X-ray band. Recent detections of a number of neutron stars with ROSAT
(Becker 1995) show good promise of enabling a study of neutron star
cooling. While spectra are still generally fit with blackbodies at
present, it has been shown (Romani 1987) that radiation transfer
through the surface layers can have very significant effects on the
emergent spectrum, depending on the composition.

Conditions on the neutron star surface are very uncertain. While the
supernova explosion has a mass cut in the iron layer, envelope
fallback, accretion during neutron star evolution, and possibly
spallation by energetic magnetospheric particles can introduce light
elements ({\it eg.}  H and He) to the neutron star surface. Because
gravitational sedimentation is rapid, the lightest element present
should dominate the photosphere and the emergent spectrum. This
uncertainty makes it important to compute emergent spectra for both
light element and heavy element atmospheres. As shown by Romani (1987)
the general trend is for light elements, with Kramer's law opacities
at high E, to have emergent spectra with large excesses in the Wien
tail over the equivalent blackbody. Heavy element atmospheres give
spectra much closer to blackbodies, but show strong line and edge
features. Recent updates of low B model atmosphere computations
(Rajagopal \& Romani 1996 (RR96); Zavlin, Pavlov \& Shibanov 1997
(ZPS96)) provide a good set of models for comparison with low field
neutron stars. As an example, RR96 found that H spectra fit ROSAT PSPC
observations of the low-B millisecond pulsar J0437--4715 much better
than did heavy element atmosphere spectra.

Most observed neutron stars, however, have surface fields at least as
strong as the surface dipole, $10^{12.5}$G or higher. Detailed
treatment of the atomic cross sections and the equation of state has
allowed Pavlov and colleagues (Pavlov et al. 1995; Shibanov et
al. 1992) to study emergent spectra from magnetized H atmosphere
neutron stars.  Unlike the recycled pulsars, though, the young, high
field neutron stars have presumably experienced little accretion. It
is therefore important to estimate the spectral effects of heavy
element atmospheres to compare with the magnetic H results. A first
attempt was pursued by Miller (1992) who obtained approximate energy
levels and cross sections for heavy atoms in strong fields (Miller \&
Neuhauser 1991) and computed emergent spectra. These opacities were
very approximate, employing only bound-free cross sections and
averaging the polarization modes. Pavlov et al. (1995) argue that
these approximations are too severe for the high field models.

In this paper we present substantially improved magnetic Fe
atmospheres, incorporating more detailed opacity and equation of state
estimates, normal mode radiative transfer and more realistic magnetic
geometries.  These approximate atmospheres will provide a useful
baseline for comparison with the magnetic H results. Since fully
detailed Fe atmospheres will be very difficult to construct, we focus
on the broad spectral differences between models and note the areas
where more detailed models can improve the spectral estimates.

\section{Atoms in a High Magnetic Field}\label{atoms}

When the electron cyclotron radius $\hat{\rho}= 2.5\times
10^{-10}(B/10^{12}\ {\rm G})^{-1/2}$ cm becomes smaller than the
approximate zero-field outermost electron orbital sizes of an ion $a
\approx a_0(Z_{\rm eff}+1)^{-1}$, ($a_0=5.2\times 10^{-9}$ cm, 
$Z_{\rm eff}$ the net charge of the ion), the ion's atomic structure
perpendicular to the field is altered significantly.  For hydrogenic
atoms of nuclear charge $Z_{\rm eff}+1$ in the ground state, this
occurs at a magnetic field $B=(Z_{\rm eff}+1)^2 B_c$, where $B_c =
2.35\times 10^9$ G.  Well below this field the electron wave functions
are approximately spherical, whereas far above it they are best
represented as cylinders given by $\psi_{m\nu}(z,\rho,\phi) =
f_{nm\nu}(z)\Phi_{nm}(\rho,\phi)$, where $(\rho,\phi)$ are cylindrical
co-ordinates about the magnetic field direction ${\hat z}$,
$\Phi_{nm}(\rho,\phi)$ are the cylindrical Landau states, and
$f_{nm\nu}(z)$ is the wave function in the direction of the field.
This assumption of separability in cylindrical co-ordinates is
sometimes called the `adiabatic approximation'.  Here $\nu$ and $m$
are the $z$ and $\phi$ quantum numbers, and $n$ is the Landau level.
The $\Phi_{0m}(\rho,\phi)$ wave functions are annular, with radius
$a_{\perp} =
\sqrt{2m+1}\hat{\rho}$ and width $\sim \hat{\rho}$ (e.g Meszaros 1992).

To compute energies and electron configurations for iron in strong
magnetic fields, we use a multiconfigurational Hartree-Fock code
developed by Neuhauser, Langanke, \& Koonin (1986; see also Miller \&
Neuhauser 1991).  This code computes the wave functions and state
energies in the `adiabatic approximation', with Landau level $n=0$
only, since the first Landau energy (cyclotron energy, $E_c = 11.6
B_{\rm 12}$keV for a magnetic field of $10^{12} B_{\rm 12}$G) greatly
exceeds both $kT$ and all the photon energies $E_\gamma$ we consider.
For the innermost electron of iron at the magnetic fields we consider,
$B=10^{12.5}$ G and $B=10^{13}$ G, this code gives binding energies
that can be low by as much as 15\% because the Coulomb field of the
nucleus is too strong for the separability assumption.  However, for
electrons with binding energies less than $\sim$3 keV, whose
transitions are most important for the observed spectrum, the code
gives individual orbital energies accurate to better than 1\%.  For
each ionization state, we have computed the energies and electron
configurations of the ground state, and the first several ($\sim 10$)
excited states which generally extend to $>5 kT$ for the typical local
atmosphere conditions.  As electrons are added to the bare nucleus, in
the ground state they occupy the ``tightly-bound'' $\nu = 0$ states,
filling $m=0,1,2,...$ in order.  These states have no $\hat{z}$ node,
and hence are localized nearer the nucleus than the ``loosely-bound''
$\nu > 0$ states.  Ground states with many electrons also fill the
$\nu=1,\;m=0,1$ orbitals, with only one electron per orbital (a spin
flip requires $\delta E
\sim E_c$).  The excited states we have are all formed by promoting
the outermost electrons to higher $m$, which costs little energy
because of the weak $m$ dependence of $a_\perp$ (above).  We use these
configurations to calculate the bound-free absorption cross section in
each of the $(+,\;-,\;z)$ polarization modes (described in \S
\ref{modes}) as a function of frequency for each individual electron,
for at least the lowest five energy states of each ion.  We also
calculate the energies of the most important configurations accessible
from the ground state via dipole allowed transitions, and some
characteristic oscillator strengths for these, for transition energies
up to 10 keV.  These configurations are made by promoting any one of
the outermost (large-$m$) electrons to $\nu > 0$.

\section{Ionization Equilibrium and the Equation of State}\label{eos}

Let the $r^{th}$-ionized atom ($Z_{\rm eff}=r$) have excited states $k$ with
energy levels $E_{r,k}$. The ionization energies are $\chi_r \equiv
E_{r,0}-E_{(r-1),0}$, while the excitation energies are
$\epsilon_{r,k} \equiv E_{r,k}-E_{r,0}$.  Then ionization equilibrium
in a strong magnetic field gives the ion densities $n_r$ as
\begin{equation}\label{kherson}
n_r = 2 \frac{n_{(r-1)}}{n_e} \left( \frac{m_e kT}{2\pi\hbar^2} \right)^{3/2}
\frac{\eta_r}{\sinh\eta_r} \frac{\sinh\eta_{(r-1)}}{\eta_{(r-1)}} 
\frac{\eta_e}{\tanh\eta_e} e^{-\chi_r/kT} 
\frac{\sum_{k=0}^{\infty}e^{-\epsilon_{r,k}/kT}}
{\sum_{k=0}^{\infty}e^{\epsilon_{(r-1),k}/kT}}\;\;,
\end{equation}
(Khersonskii 1987), where the sums are over the excitation states of
the individual ions. Here $\eta_e \equiv \hbar\omega_B/2kT$ and
$\eta_r\equiv\hbar\omega_r/2kT$, where $\omega_B=eB/m_ec$ is the
electron cyclotron frequency, $\omega_r = reB/m_{Fe}c$ is the
cyclotron frequency of the $r^{th}$-ionized atom, $m_e$ is the mass of
the electron, and $m_{Fe}$ is the mass of an iron nucleus. The sums
are the partition functions for each ion, and the relative abundance
of each excitation state is given simply by its Boltzmann factor
divided by that sum.  At high densities in our atmospheres, the
electron continuum energy can be significantly depressed; we approximate this
depression by decreasing the ionization energies by $E_{\rm dep} =
(3/5)e^2 (4\pi\rho/3m_{Fe})^{1/3}(\langle Z_{\rm eff}^2
\rangle/\langle Z_{\rm eff} \rangle) = 11.8 \rho^{1/3} (\langle 
Z_{\rm eff}^2 \rangle/\langle Z_{\rm eff} \rangle)$ eV, with $\rho$
the atmosphere density in ${\rm g/cm}^3$ (Cox \& Giuli 1968).  Since
we do not have an exhaustive list of $\epsilon_{r,k}$ to compute the
partition function sums in (\ref{kherson}), we take a constant level
spacing from the highest available excited states to complete the
partition sum as a geometric series.  This method does not include the
loosely-bound states $\nu = 1,2,3...$, of which we estimate there are
$\sim 10$ for each $m$, spread between $\sim 1/3$ the tight binding
energy and the suppressed continuum.  At the highest temperatures
present in our atmospheres, the Boltzmann suppression for these states
only partly compensates for their large number, and they may
contribute significantly to the partition function.  Under these
conditions, more detailed treatments of ionization equilibrium will
require exhaustive line lists, and a means of calculating pressure
effects on partition sums and ionization tailored to the magnetic
plasma.

Our equation of state under these extreme conditions is of necessity
approximate, despite some justified simplifying assumptions.  Although
the gas deep in our atmospheres would be degenerate at low fields,
with high $B$ the degeneracy is strongly suppressed. In particular,
the gas is non-degenerate if $T \gg T_{F} = 0.6 (\rho(Z_{\rm
eff}/26))^2/B_{12}^2$ (Hernquist 1984).  For our atmospheres,
$B_{12}^2 = 10-100$ and $\rho(Z_{\rm eff}/26) <10^3$, so the gas is
far from degeneracy throughout (Fig. 2). Also $kT \ll\hbar\omega_B$;
in the adiabatic limit, this ensures that only one Landau level is
occupied, although more exact treatments require the consideration of
higher $n$ (Potehkin, Pavlov and Ventura 1997).  The plasma parameter
$\Gamma=Z_{\rm eff}^2e^2 (4\pi n_i/3)^{1/3}/kT$, where $n_i$ is the
ion number density, is far below the solidification value ($\Gamma
\sim 170$). While our atmospheres do extend well past $\Gamma = 1$ at
high $T_{eff}$, we ignore the increasing non-ideality under these
conditions ($eg.$ Fehr and Kraft 1995).  Accordingly, we solve for the
ionization equilibrium conditions iteratively, using (\ref{kherson}),
to find the number density of all species (and hence $\rho(P,T)$)
required by the ideal gas law to produce the pressure $P(\tau)$ needed
for hydrostatic equilibrium (\S \ref{converging}).  The mean
ionization level $\langle Z_{\rm eff} \rangle$ is shown for selected
$\tau$ in Fig. 2.  We have tested the sensitivity of our final results
to the omission of non-ideality corrections to the equation of state
in two ways. First, we have adopted 20\% of the Debye-H\"{u}ckel
correction as approximately appropriate for the densities near the
base of our atmospheres (Rogers 1981), to estimate the structure
perturbation in our $10^6$K atmosphere model. Holding the temperature
run fixed, we find that although there are $\sim 5$\% variations in
density at the largest depths considered, at the surface the
differences in the emergent spectrum are below $\sim 1$\% near the
peak, and the total flux difference is $\sim 0.25$\%. We have also
made the extreme assumption that the plasma becomes an incompressible
liquid at $\Gamma=3$ and tested the effect on the emergent spectrum:
$\sim 5\%$ changes were seen in the emergent flux near the weakly
bound lines, but significant continuum variation was found only for
$E_\gamma > 20 kT_{\rm eff}$.  We therefore ignore plasma corrections
in the following computations.

\section{Opacities}

\subsection{Radiation Propagation in the Magnetized Plasma}\label{modes}

In a magnetized plasma, radiation may propagate only in two specific
polarization modes, called the normal modes.  Their properties are
obtained in general by finding the polarizability tensor of the
plasma, and using it to solve Maxwell's equations for the case of a
plane wave.  For a collisionless magnetized plasma, the resulting mode
polarization components are easily given (Ginzburg 1970) in the
co-ordinate system $(x',y',z')$, where $z'$ is in the propagation
direction $\hat{k}$, and the magnetic field direction $\hat{B}$ is in
the $y'$-$z'$ plane, at angle $\theta$ from the $z'$ axis.  The
relative strengths of the components for modes $j=1,2$ are
\begin{equation}\label{ginzmodes}
\begin{split}
\frac{e_{y'}^j}{e_{x'}^j} = -i\frac{2u^{1/2}(1-v)\cos\theta}
{u\sin^2\theta + (-1)^{j-1} (u^2\sin^4\theta + 4u(1-v)^2\cos^2\theta)^{1/2}} \\
e_{z'}^j = -i\frac{u^{1/2}v\sin\theta}{u-(1-v)-uv\cos^2\theta}e_{x'}^j + 
\frac{uv\cos\theta\sin\theta}{u-(1-v)-uv\cos^2\theta}e_{y'}^j\;\;,
\end{split}
\end{equation}
where $u\equiv(\omega_B/\omega)^2$, and $v\equiv(\omega_p/\omega)^2$
with $\omega_p = 4\pi n_e e^2/m$ the plasma frequency.

We normalize these components, and rotate them to the basis $(x,y,z)$
with $z$ in the $\hat{B}$ direction, and $\hat{k}$ in the $y$-$z$
plane, retaining the definition of $\theta$, not to be confused with
the angle $\theta_n$ between $\hat{k}$ and the atmosphere's outward
normal.  Finally we construct $e_+^j \equiv (e_x^j + ie_y^j)/\sqrt{2}$
and $e_-^j \equiv (e_x^j - ie_y^j)/\sqrt{2}$, the polarization
components in the rotating co-ordinate basis in which the opacities
are most easily calculated.  The indices of refraction are
\begin{equation}\label{nref}
n_j^2 = 1 - \frac{2v(1-v)}{2(1-v) - u\sin^2\theta + 
(-1)^j (u^2\sin^4\theta + 4u(1-v)^2\cos^2\theta)^{1/2}}
\end{equation} 
For the large $u$ we consider, the $j=2$ mode has an imaginary index
of refraction ($n=i\,{\widetilde n}$, $\widetilde n$ real) below the 
plasma frequency, when
$1<v<\sec^2\theta$.  In this case we add the extra ``skin-depth''
opacity $\tau_\delta(\omega) = {\widetilde n}\omega/c$ to the total 
mode opacity of \S \ref{totradop}. In practice all but a negligible
fraction of the emergent flux from our atmospheres is above the plasma
frequency at the depth of spectrum formation.

In fact, absorptive terms in the polarizability tensor corresponding
to the various opacities affect the mode decomposition.  As a test, we
treat electron-ion collisions as in Ginzburg (1970), approximating
non-magnetic free-free absorption: the effect is negligible except
well below the plasma frequency.  Nonetheless we expect that proper
inclusion of the atomic opacities will change the mode polarizations
somewhat, especially in the line regions.  To date, the only treatment of
this effect is for hydrogen (Bulik \& Pavlov 1996).

\subsection{Thomson Scattering and Free-Free Absorption Opacities} 

Meszaros (1992) gives the normal mode electron scattering and
free-free opacities as
\begin{equation}
\sigma = (\sigma_T,\sigma_{ff}) \left[ \frac{\omega^2}{(\omega+\omega_B)^2} 
|e_+|^2 g_\perp + \frac{\omega^2}{(\omega-\omega_B)^2}|e_-|^2 g_\perp
 + |e_z|^2 g_\parallel \right]
\end{equation}
For Thomson scattering, $\sigma_T = (8\pi/3)\alpha_F^2 (\hbar/mc)^2$
is just the non-magnetic cross-section, where $\alpha_F$ is the fine
structure constant, and $g\equiv 1$.  For free-free absorption
(Brehmsstrahlung),
\begin{equation}
\sigma_{ff} = 4\pi^2 \alpha_F^3 \frac{\hbar^2 c^2 }{m}  
\langle Z_{\rm eff}^2 \rangle n_i
\frac{1-\exp(-\hbar\omega/kT)}{\omega^3 (\pi mkT/2)^{1/2}}\;\;,
\end{equation}
is also the field-free cross-section, and the Gaunt factors are given
by
\begin{equation}\label{gauntfac}
\begin{split}
g_\perp(\omega,T,B) = \frac{1}{2} \int_{-\infty}^\infty \exp[-p^2/(2mkT)]
\frac{C_1(a_+) + C_1(a_-)}{(p^2 + 2m\hbar\omega)^{1/2}} \\
g_\parallel(\omega,T,B) = \int_{-\infty}^\infty \exp[-p^2/(2mkT)]
\frac{a_+ C_0(a_+) + a_- C_0(a_-)}{(p^2 + 2m\hbar\omega)^{1/2}} \\
a_\pm = ( p \pm [p^2 + 2m\hbar\omega]^{1/2})^2 (2m\hbar\omega_c)^{-1} \\
C_0(a) = \exp(a) E_2(a)/a \;\;\;\;\;\;\;\;\;\;\; 
C_1(a) = \exp(a) [E_1(a) - E_2(a)]\;\;,
\end{split}
\end{equation}
(e.g. Meszaros 1992), where $E_1$ and $E_2$ are the exponential
integral functions.  Under the conditions of our atmospheres, these
Gaunt factors are $\sim$1 to 10; they slightly steepen the inverse
dependence of opacity on frequency.

\subsection{Bound-Free and Bound-Bound Opacities; Line Broadening}
\label{bfbb}

Taking the bound-free $(+,\;-,\;z)$ basis cross sections from the
Hartree-Fock wave function sums, we add the contributions from all
electrons in each ion.  The result for any excitation state of an ion
is a bound-free edge and a falloff with complex structure.  In the
case of hydrogen, these features can be affected by the thermal motion
of the ion transverse to the field (Kopidakis, Ventura \& Herold,
1996); and, if adiabaticity is not assumed, by resonances with states
with non-zero Landau level (Potekhin, Pavlov \& Ventura, 1997).  We
expect the former effect to be much diminished for iron, while the
latter could be as significant as other inaccuracies inherent in the
`adiabatic approximation' (see \S \ref{atoms}).  When atoms are
excited to a state for which an explicit bound-free cross section is
not available, we assign them the cross section of the highest
available state.

To estimate the line opacity we use a list of energy levels from the
Hartree-Fock modeling. Because of uncertainty in the energy level
computations, line energies may be in error by as much as 10\%, but
the general distribution of bound-bound transitions is representative
of the true spectrum. We focus here on dipole allowed transitions,
which obey the selection rules $\Delta m = \pm 1$ for even $\Delta
\nu$, and $\Delta m = 0$ for odd $\Delta \nu$ (Ruder et al. 1994).  We
find that in the adiabatic approximation, the energy level spectra for
individual ions are roughly self-similar.  Therefore, for each
ionization state we use the $\Delta m = 0$, $\Delta \nu = 1$
transition of the outermost ground state electron, which we always
have, to (linearly) re-scale the most extensive line energy lists
available (for $Z_{\rm eff}=2$ at the lower field, $Z_{\rm eff}=0$ at
the higher) and apply it to the ion.  For most ions we have explicitly
computed a few allowed transition energies; these are always used, but
in any case they match well to the predicted energy from the re-scaled
line lists.  We have attempted to compute a `complete' set of dipole
allowed transitions from the ground states only; we generally do not
have explicit energies for all relevant bb transitions from excited
states. To approximate the total distribution of oscillator strength
from these transitions, we estimate the number of missing lines for
each excited state and assign the ground-state line spectrum to the
unaccounted-for fraction.  While we have not computed the oscillator
strengths for all transitions in detail, we can estimate the line
opacity by noting that the strengths, defined by $\int\sigma(E){\rm
d}E = (2\pi^2e^2\hbar/mc) f_{\rm osc}$, take on characteristic sizes
for the different transitions.  Furthermore, the type of transition
determines to which component of the polarization basis the line
couples.  We use $f_{\rm osc}($couples to$\;+\;{\rm mode}) \approx
0.06$ ($\Delta m=1,\;\Delta\nu=0$); $f_{\rm osc}(z\;{\rm mode})
\approx 0.8,0.06$ ($\Delta m=0,\;\Delta\nu=1,3$); and $f_{\rm
osc}(+,-) \approx 5\times 10^{-4}, 5\times 10^{-5}$ ($\Delta
m=1,-1;\;\Delta\nu=2$) for the oscillator strengths at
$B=10^{12.5}$G. For $B=10^{13}$G, the corresponding strengths are
$f_{\rm osc}(+) \approx 0.04$ ($\Delta m=1,\;\Delta\nu=0$); $f_{\rm
osc}(z) \approx 0.4,0.03$ ($\Delta m=0,\;\Delta\nu=1,3$); and $f_{\rm
osc}(+,-) \approx 9\times 10^{-5}, 9\times 10^{-6}$ ($\Delta
m=1,-1;\;\Delta\nu=2$).  These oscillator strengths are accurate to
within a factor of $\sim 2$, which is sufficient for our modeling
needs.

We include two very different broadening mechanisms in our
calculation.  First, at the high densities and temperatures
characteristic of neutron star atmospheres, line features experience
significant collisional broadening.  For an electron in an orbital of
characteristic radius $a$ about a nucleus of charge Z, the energy
shift due to the dipole induced by an external electric field $F$ is
$\Delta E \sim F^2a^3/Z$.  For collisional impact, we have $F =
e/r_{\rm impact}^2$, and $Z=Z_{\rm line}$ the net charge interior to
the electron absorbing the photon.  We use the Weisskopf method to
approximate the broadening, since the impinging electrons are confined
by the magnetic field to move in a straight line, giving a Lorentzian
cross-section with width
\begin{equation}\label{collwid}
\hbar\Gamma_{\rm coll} = \pi^{5/3} 2^{1/3} \hbar^{1/3} a^2\;e^{4/3} 
(\frac{kT}{m_e})^{1/6}\frac{n_e}{Z_{\rm line}^{2/3}}\;\;\; 
\sim\;\; 41.5\;T_6^{1/6} \frac{(n_e/10^{24})}{Z_{\rm line}^{8/3}}
\left[ \frac{a}{(a_0/Z_{\rm line})}\right]^2\; {\rm eV}\;\;,
\end{equation}
which comes from the amplitude and $r$-dependence of $\Delta E$ (see
Mihalas 1978).  For $a$ we use the largest dimension of initial or
final state, with dimension $a_\perp$ transverse to the field as given
in \S \ref{atoms}, and $a_z$ parallel to the field estimated as
follows.  Orbitals with $\nu=0$ are localized near the nucleus; we
need their $a_z$ only if it is bigger than $a_\perp$, in which case it
is given, through minimizing the total energy of an annular charge
distribution, by $a_z = a_0/\ln(a_0/a_\perp)$ (e.g. Meszaros 1992 \S
2.4).  For $\nu > 0$ we estimate $a_z = (a_0/Z_{\rm line})
[(\nu+1)/2]^2$, which comes analytically from the Schr\"{o}dinger
equation for $f_{0m\nu}$ (\S \ref{atoms}) for odd $\nu$ states of
hydrogen at high field (Ruder et al. 1994, \S6.4.4).  Since energies
vary smoothly with $\nu$ regardless of parity, both in that work and
in ours, we use the above estimate for even $\nu$ also.

Secondly, the thermal motion of the ions past the large $B$ also causes 
significant energy level
perturbations (Pavlov \& Meszaros 1993). We can approximate this
effect by a Lorentz electric
field $F = (v_\perp/c) B$, where $v_\perp$ is the velocity transverse
to the magnetic field.  As in the case of hydrogen, 
we take this `magnetic broadening' to be one-sided,
since the Lorentz field always results in
elongation of the otherwise circular transverse wave function.  We
estimate the energy shift as above by $\Delta E
\sim F^2a^3/Z_{\rm line}$, taking $a$ to be the larger $a_\perp$ of
initial and final state, since $F$ must be transverse to $B$.  Using
$v_\perp \sim \sqrt{2kT/m_{\rm Fe}}$ we get
$F=5.8\times10^7 B_{\rm 12} \sqrt{T_6}$ cgs, yielding a width of
\begin{equation}\label{magwid}
\hbar\Gamma_{\rm mag} = 0.035 (2m+1)^{3/2}\; \frac{T_6 \sqrt{B_{\rm 12}}}
{Z_{\rm line}}\;\;\;{\rm eV}\;\;.
\end{equation}
Making proper allowance for the higher thermal velocity of hydrogen,
equation (\ref{magwid}) agrees with the numerical example in Pavlov \&
Meszaros (1993) within a factor of 2.  Note that since it is
independent of density, magnetic broadening is most important near the
surface of our atmospheres.  We add the two broadening mechanisms,
giving a redward half-Lorentzian with width $\hbar\Gamma_{\rm R} =
\hbar\Gamma_{\rm mag} + \hbar\Gamma_{\rm
coll}$, and a blue line wing with width $\hbar\Gamma_{\rm B} =
\hbar\Gamma_{\rm coll}$.  The line profile, normalized to the $f_{\rm
osc}$ as above, is thus
\begin{equation}\label{linered}
\sigma(E){\rm d}E = \frac{\pi e^2\hbar f_{\rm osc}}{mc} 
\left( \frac{ 2\Gamma_{\rm R} }{ \Gamma_{\rm R} + \Gamma_{\rm B} } \right)
\left( \frac{\hbar\Gamma_{\rm R}}{(E - E_0)^2 + (\hbar\Gamma_{\rm R}/2)^2} 
\right) {\rm d}E\;\;\;\;\;\;\;\;\;\;
E \le E_0 
\end{equation}
\begin{equation}\label{lineblue}
\sigma(E){\rm d}E = \frac{\pi e^2\hbar f_{\rm osc}}{mc} 
\left( \frac{ 2\Gamma_{\rm R} }{ \Gamma_{\rm R} + \Gamma_{\rm B} } \right)
\left(\frac{(\Gamma_{\rm B}/\Gamma_{\rm R}) \hbar\Gamma_{\rm B}} 
{(E - E_0)^2 + (\hbar\Gamma_{\rm B}/2)^2} 
\right) {\rm d}E\;\;\;\;\;\;\;\;\;\;
E \ge E_0\;\;\; ,
\end{equation}
where $E_0$ is the line energy.  As this net broadening is sufficient
to blend many of the lines, the exact values of the line energies are
not crucial in estimating the b-b contribution to the overall opacity.

\subsection{Total Radiative Opacity}\label{totradop}

When our atmosphere calculation calls for the radiative opacity in
each mode at a given angle and frequency, for some ambient temperature
and density, we sum the scattering, free-free, bound-free and
bound-bound $(+,\;-,\;z)$ cross sections for that frequency over all
ionization and excitation states weighted by (\ref{kherson}) at the
appropriate temperature and density, giving
$\sigma_{+,\;-,\;z}(\omega,\theta,T,\rho)$. The mode cross-sections are
then given by
\begin{equation}\label{basis}
\sigma_j(\omega,\theta,T,\rho) = \sum_{i=+\;,-,\;z} e^j_i(\omega,\theta)^2
\sigma_i(\omega,T,\rho) \qquad\qquad j=1,2
\end{equation}
We convert these from cm$^2$/particle to cm$^2$/g, and multiply by the
density to get the opacity $\kappa_j(\omega,\theta,T,\rho)$ in
cm$^{-1}$.  

\begin{figure}[htb]
\plotone{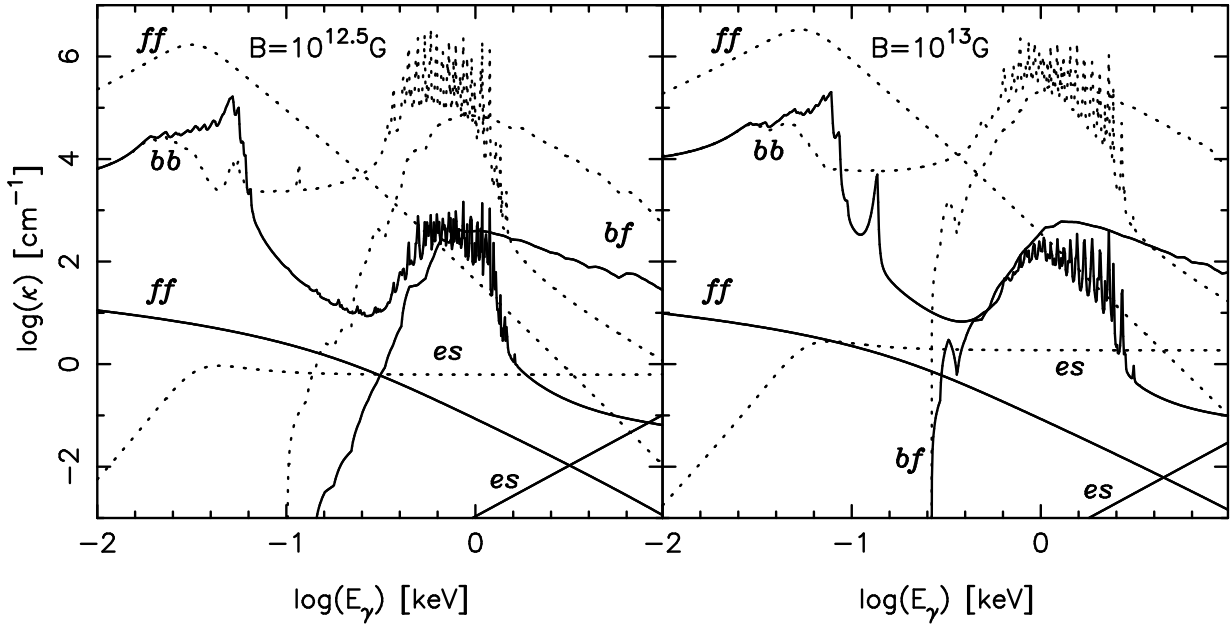}
\caption{Left: Contributions to the radiative opacity at $\theta = 
55^\circ$, at total mean optical depth (defined in \S
\ref{converging}) $\tau_T=1$ in our $B=10^{12.5}$G, $T_{\rm eff} =
10^6$K, $\Theta_B = 0^\circ$ Fe atmosphere. The two modes are shown by
solid and dashed lines respectively. Note that significant b-f opacity
is present at large $E$.  b-b transitions between tightly bound states
appear at low energy; tight-weak transitions are at higher energy.
Right: Same for $B=10^{13}$G.}
\end{figure}

To illustrate the relative importance of different contributions to
the opacity, we show opacity spectra for the two normal modes at
$55^\circ$ to the magnetic field, under characteristic atmosphere
conditions.  Free-free opacity dominates at low energies, but b-f and
b-b processes contribute significant opacity at higher energies;
electron scattering is never significant in these Fe atmospheres.

\section{Energy Transfer in the Atmosphere}

\subsection{Radiative Transfer}

The magnetic field, at an angle $\Theta_B$ to the atmosphere normal,
causes significant anisotropies in the energy transfer. In addition,
ZPS96 have shown that even without a field there can be substantial
limb darkening and that the emergent spectrum will vary with $\mu =
\cos \theta_n$ the cosine of the viewing angle to the atmosphere
normal. For emission from a small region on the star, these effects
will cause substantial variation with the viewing angle. We therefore
follow the angular dependence of both internal radiation field and
emergent spectrum, although our atmosphere temperature structure
remains plane parallel.

We define a grid of angular co-ordinates $(\theta,\phi)$ about the
magnetic field direction.  Given the state of the atmosphere at each
depth, we use the opacities of \S 3 to compute frequency- and
$\theta$- dependent optical depths, by summing
$\tau_j(d,\omega,\theta) = \int \kappa_j (\omega,\theta,T,\rho) {\rm
d}z$ from the surface to atmosphere layer number $d$, for each theta
in our grid.  This precedes any consideration of the atmosphere normal
direction, and the sums are done using the normal thickness ${\rm d}z$
of each layer.

We next compute radiation fields at each $\theta$ and $\phi$ at a
given atmosphere level.  The intensity of outward going radiation is
given by
\begin{equation}\label{I_trans_out}
I_j[\tau_j(\omega,\theta),\mu] = \int^\infty_{\tau_j} [B_\omega(\tau')/2] 
e^{-(\tau' - \tau_j)/\mu} {\rm d}\tau'/\mu \;\;\;\;\;\;\;\;\; (0 \le \mu \le 1)
\end{equation}
and inward going radiation by
\begin{equation}\label{I_trans_in}
I_j[\tau_j(\omega,\theta),\mu] = \int^{\tau_j}_0 [B_\omega(\tau')/2]
e^{-(\tau_j - \tau')/(-\mu)} {\rm d}\tau'/(-\mu)\;\;\;\;\;\;\;\;\;(-1 \le \mu \le 0)
\end{equation}
where $\mu=\cos\theta\cos\Theta_B + \sin\theta\sin\Theta_B\cos\phi$
(taking the normal direction to have $\phi=0$ by definition) and
$B_\omega(T) = (\hbar\omega^3/[2\pi^2c^2 (e^{\hbar\omega/kT} - 1)]$ is
the Planck function, half of which gives the source function in each
mode for LTE.  When $\Delta \tau_j(\omega,\theta) > 20$ in a given
layer we use the diffusion approximation for that frequency and mode:
$I[\tau_j(\omega,\theta),\mu] = B_\omega[\tau_j(\omega,\theta)] + \mu
\partial B_\omega /\partial
\tau_j(\omega,\theta)$. The intensities are then summed to find the
astrophysical flux $F_{j\omega} = (1/\pi) \iint \mu
I_j(\omega,\theta,\phi){\rm d}\Omega$. We find that 18 zones in
$\theta$ and 9 zones in $\phi$ (zero to $\pi$ only, by symmetry) give
better than 0.1\% accuracy in the intensity integral even for test
functions with rapid variations in $\theta$ and $\phi$. To monitor
atmosphere flux the above computations are performed on 200 energy
bins, logarithmically spaced between 1eV and 10keV.

\subsection{Electron Thermal Conductivity}

Electron thermal conduction provides a significant component of the
heat flux at large depth in our coolest atmosphere models. The
electron thermal conductivity is significantly modified under high
field conditions (Hernquist 1985). Because the electron gas in our
atmospheres is non-degenerate (see \S \ref{eos}), and occupies only
the lowest Landau level, the conductivity $\lambda_\parallel$
(Hernquist's $\kappa_\parallel$) along ${\hat z}$ can be readily
computed by evaluating the integrals in Hernquist's Eq (3.5), using
(3.8) and (3.10).  The chemical potential is given by (2.12), and its
derivative required by (3.10c) is taken analytically.  Electron
conduction perpendicular to $B$ will be much smaller, and we take it
to be zero.  We find that in our coldest $B=10^{13}$G atmosphere, in
which it is most significant, the electron conductivity carries up to
80\% of the heat flux at $\tau_{\rm rad}=10$ (defined below).  The
extra mode of transport affects the temperature structures and
emergent spectra significantly.

\section{Converging the Atmosphere and Obtaining Spectra}\label{converging}

Deep within the atmosphere, the optical depths are large for {\it all}
frequency bins, and radiation flows in the diffusion limit with flux
$F_\omega = (1/\pi)\iint \mu(\theta,\phi)^2({\rm d}B_\omega/{\rm
d}\tau_{\omega\theta})\sin\theta {\rm d}\theta{\rm d} \phi$.  We
therefore define a radiative mean opacity
\begin{equation}\label{radmean}
\kappa_{rad}^{-1} = \frac{\pi}{4\sigma T^3} \left[\int_0^{\infty}
\frac{\partial B_{\omega}}{\partial T} \frac{3}{4\pi}\left( \int_0^{2\pi}
\int_0^{\pi} \frac{1}{2}\sum_j \frac{\mu(\theta,\phi)^2}{\kappa_j(\omega,\theta)}
\sin\theta{\rm d}\theta{\rm d}\phi \right) {\rm d}\omega \right]
\end{equation}
equivalent to the Rosseland mean opacity; the corresponding optical
depth scale is $\tau{\rm rad}$.  The total mean opacity is then given
by $\kappa_{\rm T}^{-1} = \kappa_{\rm rad}^{-1} +
\kappa_c^{-1}$, where $\kappa_c = 16 \sigma T^3/(3\lambda_\parallel{\rm cos}
\Theta_B)$ gives the effective conductive contribution. This total opacity 
defines a useful depth scale $\tau_{\rm T} < \tau{\rm rad}$, as the
temperature structure tends to the grey solution at large depth on
this scale. Our atmospheres have 146 depth zones with $-6 < \log
\tau_{\rm T} < 2 $; the very low $\tau$ layers are required by the
extreme frequency and angle dependence of our opacities.

As usual we relate the optical depth zones to physical depths by
summing the equation of hydrostatic equilibrium down from the surface:
${\rm d}P = [g_s \rho/\kappa_{\rm T}(\tau_{\rm T})]{\rm d}\tau_{\rm
T}$.  Because $\kappa_T = \kappa_T(\rho, T)$, we solve iteratively for
the physical zone size ({\it cf.} RR96). We then evaluate the total
radiative flux $F_{rad} = \sum_{j=1,2} \int F_{j\omega} {\rm d}
\omega$ as above at each depth zone, and add the conductive heat flow
$F_c = (\lambda_\parallel /\pi) {\rm d}T/{\rm d}z$ to obtain the total
energy flux at each depth.  This is adjusted to the constant flux
solution, using the Lucy-Uns$\ddot{\rm o}$ld temperature correction
scheme, as in RR96. We iterate until flux errors are $< 1$\%
throughout the atmospheres.

In the surface layers, we also consider the behavior of $f=\int{\rm
d}\omega\int {\rm sin}\theta\,{\rm d}\theta{\rm d}\phi \sum_{j=1}^2
\kappa_j (I - B_\omega /2)$. Although convergence to $f=0$ is a useful
diagnostic, we find that a simple $\Lambda$ iteration scheme using
$f$, as suggested by Shibanov et al. (1995), provides no help in
reaching convergence. Full scale Feautrier schemes could, however,
speed convergence considerably.

As noted above, both angle dependent specific angular fluxes and the
total emergent spectrum are of interest. These are computed from
equation (\ref{I_trans_out}) at the surface layer at $10^3$ log-spaced
energies for higher spectral resolution.

\section{Results and Conclusions}

We converge atmospheres with effective temperatures $T_{\rm eff} =
10^{5.5}, 10^{6.0}$ and $10^{6.5}$K, with vertical ($\Theta_B=0$)
magnetic fields of $10^{12.5}$ and $10^{13}$ Gauss.  To examine the
dependences on magnetic geometry, we also converge atmospheres with
$\Theta_B= 45^\circ$ and $90^\circ$ at $10^{6.0}$K for both field
strengths.

\begin{figure}[htb]
\plotone{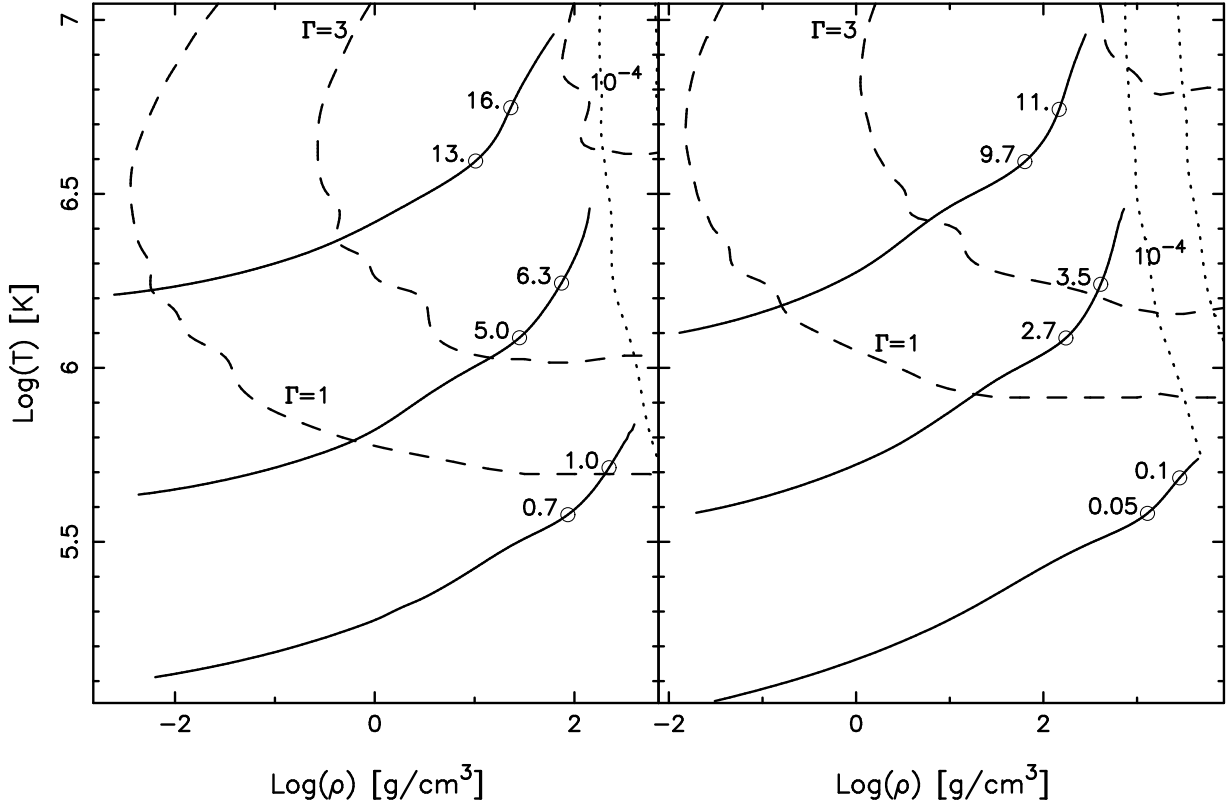}
\caption{Left: $B=10^{12.5}$G, $\Theta_B=0$ atmosphere 
structures ($\rho-T$ plane) for ${\rm log}(T_{\rm eff})=5.5,6.0,6.5$
superimposed on curves for $E_F/kT = 10^{-4}, 10^{-3}$ and
$\Gamma=1,~3$ and 10 curves. Positions of $\tau_{\rm rad}=1$ and $10$
(see \S \ref{converging}) are marked with open circles.  Labels by the
points give the mean ionization level at these depths.  Right: Same
for $B=10^{13}$G.}
\end{figure}

Figure 2 gives the temperature structure computed for the $\Theta_B=0$
atmospheres. Open circles show the locations of the mean radiative
optical depths 1 and 10, labeled with the average ionization level
$\langle Z_{\rm eff} \rangle$ at these points. The temperature
structure may be compared with background curves showing $\Gamma$ and
$E_F/kT$; the depth zones dominating the spectrum lie in the gaseous
non-degenerate regime.

Figure 3 shows the emergent spectra for the normal field atmospheres
--- these can be compared with the corresponding blackbodies. As in
Romani (1987), RR96 and ZPS96 we see that the rich absorption spectrum
of the heavy element gives an overall opacity trend much flatter than
the steep Kramer's law fall-off of H and He. This assures that the
spectra are globally closer to a blackbody shape than the light
element atmospheres.  In fact, the bound-free edges at $\sim
0.3-1.0$keV provide a large increase in opacity that gives Fe spectral
{\it deficits} above the Wien peak.  Qualitatively, this is similar to
the effect of L-edge absorptions in non-magnetic iron (RR96). Looking
at the opacities and at the absorption features in the spectra, one
can see that the magnetic Fe transitions are divided, roughly, into two
energy ranges: $\Delta \nu=0$ transitions below $\sim 100$eV (for our
lower field) and all other transitions, mostly at energies above
300eV. Interestingly, for typical neutron star magnetic fields these
transitions are at energies roughly comparable to the M- and L-shell
transitions in non-magnetic Fe. The resulting spectral shapes
therefore have some similarities to non-magnetic iron. Note that we do
not expect the detailed line positions in our atmospheres to be
accurate; the general appearance of these emergent spectra should
however be similar to more accurate results. Although more complete
atomic data should give an even richer line spectrum, we do find that
certain species dominate at depths where the lines are forming. Thus
high resolution X-ray spectra can in principal extract useful
information about the Fe atmosphere structure. In particular, the
rather strong line broadening, most visible at low E, does not prevent
discrete lines from being visible at higher energies. At low energies
lines are also visible, but equivalent widths are questionable because
of uncertainty in the detailed temperature structure in the very low
$\tau$ surface layers.

\begin{figure}[htb]
\plotone{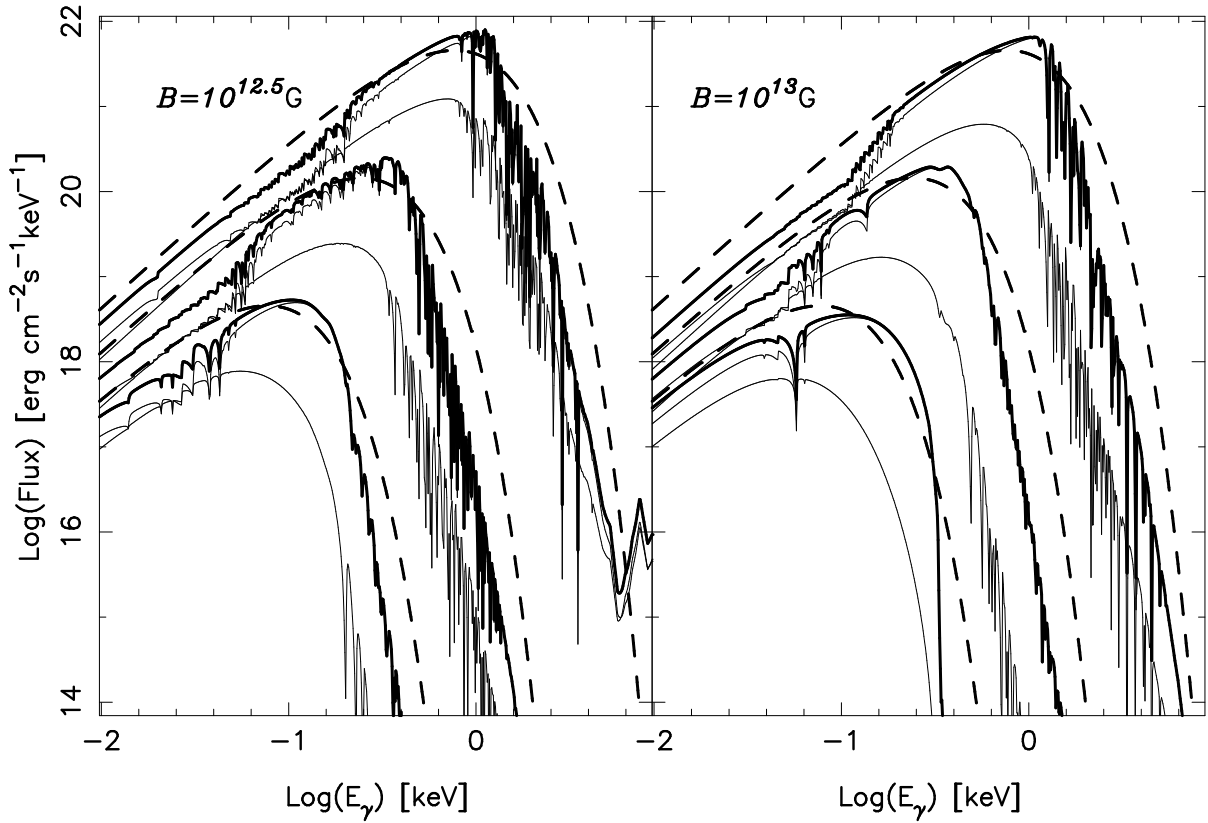}
\caption{Left: Emergent Fe spectra for $B = 10^{12.5}$G 
and $\Theta_B=0$ at ${\rm log}(T_{\rm eff})=5.5,~6.0$ and $6.5$,
showing the two modes (thin solid lines), the total flux (heavy solid
line) and corresponding black-body spectrum (dashed). Right: Emergent
spectra for $B = 10^{13}$G.}
\end{figure}

In Figure 3 we also show the contributions of the two normal modes to
the emergent flux at each temperature. While we do not discuss it in detail
here, it is clear that the thermal emission will show a strong energy
dependent polarization. This polarization also depends on the magnetic field
geometry; thermal polarization from neutron star surfaces should be
pulsed at the star's spin period.

In practice, we expect quite substantial variation of the emergent
spectrum over the surface. If due to a coherent dipole, the variation
of magnetic field strength and orientation will affect the bulk flow
of thermal flux from the interior to the star surface. For example
Shibanov and Yakovlev (1996) indicate an effective temperature
decrease as large as $\sim 7$ from the magnetic pole to the magnetic
equator for a $10^{13}$G, $\langle T_{\rm eff} \rangle=10^6$K neutron
star. Such temperature dependence implies large thermal emissivity
variation with viewing angle. For a given $T_{\rm eff}$ we find that
the broadband spectral dependence on $\Theta_B$ is quite modest (see
also Pavlov, et al. 1995). However, for a surface with varying $T_{\rm
eff}$, including the spectral dependence imposed by an atmosphere
(e.g. as modeled for Geminga by Meyer, Pavlov and Meszaros 1994) can
give a complex, energy dependent pulse behavior.  In fact, spectral
features are quite dependent on the magnetic field strength (Figure
3). At a minimum, even if the large scale dipole dominates the field
at the surface, there will be a factor of two difference in $B$
between the magnetic equator and the magnetic poles. Higher order
multipole contributions are also possible; the large field variations
will blur any spectral features significantly.

\begin{figure}[htb]
\plotone{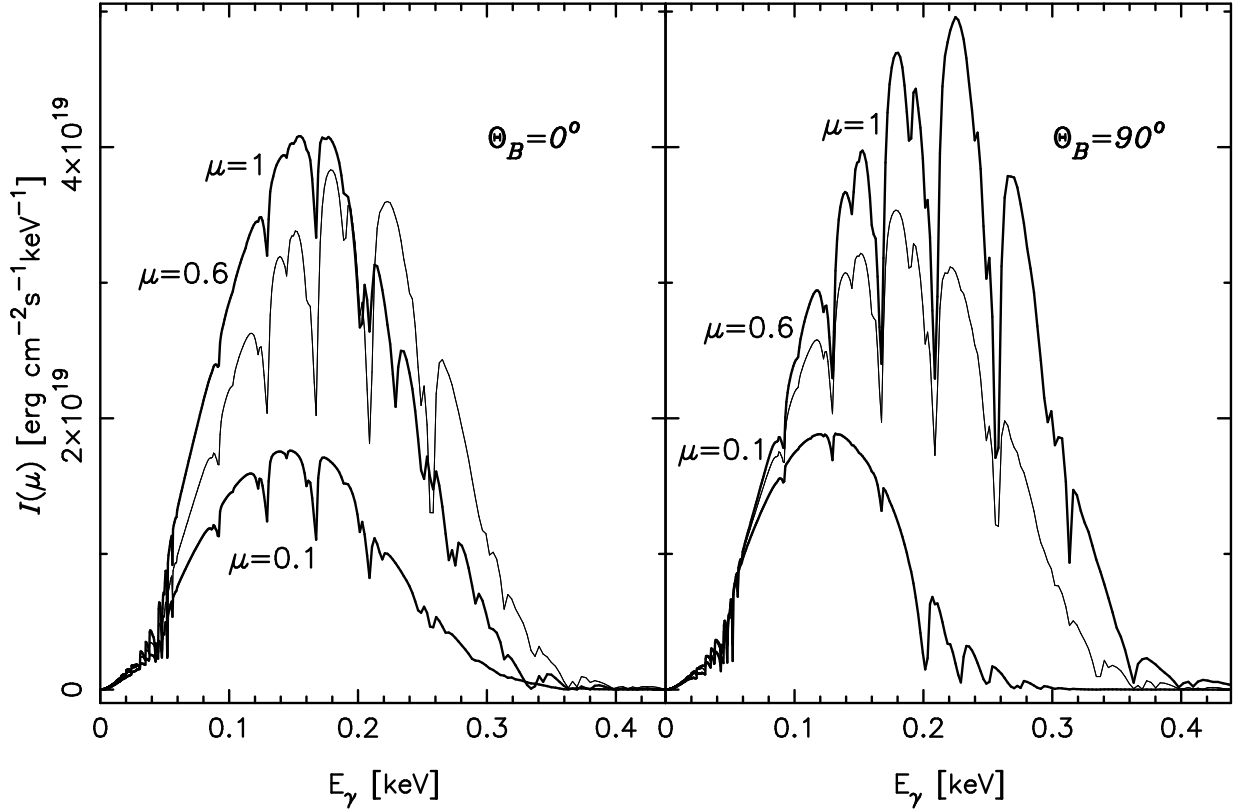}
\caption{Left: Emergent Fe atmosphere emissivity as a function of
viewing angle for $B=10^{12.5}$G, $T=10^{5.75}$K, $\Theta_B=0$. Total
fluxes are shown at $\mu = 0.1,0.6$ (lighter curve) and 1.0. Note that
the limb darkening is a strong function of energy.  Right: Viewing
angle dependence for $\Theta_B=\pi/2$. The limb darkening, spectral
features and polarization properties are a strong function of magnetic
geometry.}
\end{figure}

Finally in addition to the magnetic field variation and effective
temperature variation, there will be a significant modulation of the
neutron star thermal spectrum with changes of the observer's viewing
angle to an emitting region on the stellar surface.  ZPS96 have
emphasized the `limb darkening' decrease in the emitted flux with
increasing $\mu$ for non-magnetic atmospheres. In magnetized
atmospheres the $\mu$ dependence can be extremely complicated in
general.  One basic effect is visible below $E_\gamma= 0.1$keV
in the normal field case of Figure 4; 
here one of the two normal modes has a significant $z$
polarization component, which experiences a much higher continuum
opacity unsuppressed by the field.  The opacity increase acts in
concert with the usual limb darkening to diminish the flux at small
$\mu$.  For tangential $\Theta_B=90^\circ$, on the other hand, when
$\mu\sim 0$ in the ${\hat z}$-${\hat n}$ plane both normal modes are
primarily $(+,-)$, resulting in strongly suppressed continuum opacity.
The extra transparency counteracts the limb darkening to some extent;
the emissivity varies little below 0.1keV in the right panel of Figure 4.
Thus a simple dipole field neutron star would have a relatively
bright limb at low $E_\gamma$ when viewed along the magnetic axis. 

However, other effects can complicate this simplest picture.  First,
the lines and line wings can couple selectively to a single
polarization component, and block out the modes transparent to the
continuum.  Secondly, when $\mu \sim 1$, the normal modes are
circularly polarized and similar to one another, both containing small
amounts of the $z$ component; while for $\mu \lesssim 0.7$ the modes
are linearly polarized, and the mode perpendicular to the $B$-${\hat
k}$ plane is almost completely $z$-free.  Thus when the $z$ opacity
excess is too extreme, the $z$-free mode can have larger intensity at
small $\mu$ than both modes together at $\mu \sim 1$.  This effect is
visible above 0.2 keV in Fig. 4, where strong line wings are coupled to the
$z$ component.  In some cases even the continuum-controlled thermal
peaks are dominated by this effect; however, any small change in
normal mode composition can modify this behavior.  Clearly models
of purely thermal pulsed emission from neutron stars can be complex,
with effective temperature, line feature position and limb darkening
behavior controlled by the local magnetic geometry. These effects may
be somewhat blurred by gravitational defocusing of the emergent
radiation (Page 1995), but substantial variation in the atmosphere
features and overall emissivity are expected as the star rotates.

Our models suggest that if pulsars possess iron atmospheres, upcoming
high throughput, moderate resolution spectroscopy missions such as XMM
and AXAF will have a wealth of spectral features for diagnosis of
atmosphere temperature structure and magnetic geometry. Data providing
such rich spectra will certainly motivate more complete magnetic Fe
models.  Improvement of the atomic data sets used to generate the opacities
will be essential for such models.  Even exhaustive lists of
configurations and lines, though, will have to be coupled with a more
sophisticated treatment of the non-ideal equation of state, along the
lines of an activity expansion of the grand canonical ensemble
(Rogers 1986), or an occupation probability formalism (Hummer \&
Mihalas 1988).  Both these approaches, which have been used for
non-magnetic plasma opacity calculations by the OPAL and OP
collaborations respectively, avoid the somewhat ad hoc cutoff
procedures necessary in using equation (\ref{kherson}).  A fully
consistent opacity/EOS system will also need to include the effects of
plasma non-ideality on atomic (b-f and b-b) opacity.  This will
require calculation of states and energies of non-isolated atoms,
desirable but prohibitively difficult at present.  For the line
opacities, the dipole selection rules and simple coupling of each line
to a single polarization component are known to be a first
approximation only, subject to complication by all forms of
broadening.  Presumably more detailed atomic modeling will address
these topics, and produce more precise oscillator strengths as well.
Finally, more realistic magnetic geometries and inclusion of (weak)
mode coupling through scattering will have some effect on the emergent
flux. Nonetheless, we expect our magnetic models have captured the
qualitative behavior of heavy element atmospheres. In simplest terms
this is that the wealth of spectral features bring the overall
spectrum shape closer to a blackbody than for light element surfaces.

\begin{figure}[htb]
\plotone{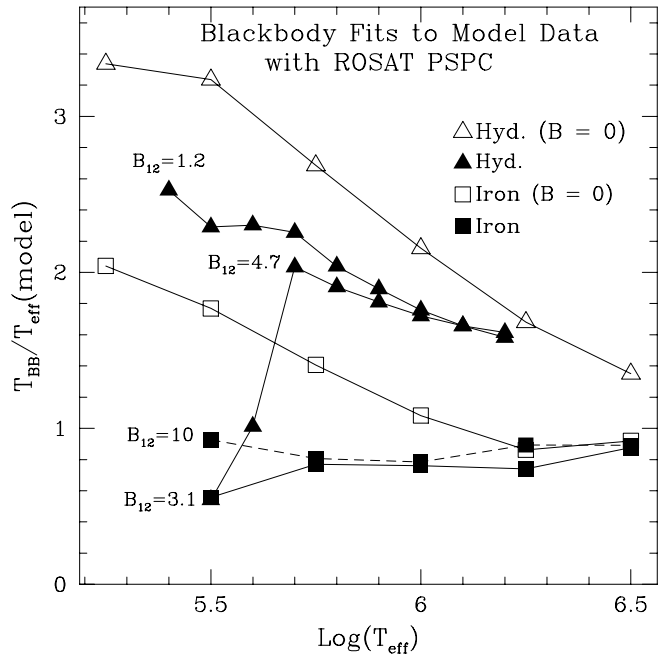}
\caption{Inferred $T_{\rm eff}$ when fitting a blackbody to a simulated
3000 count ROSAT PSPC spectrum produced by a redshifted ($z=0.306$),
absorbed ($N_H=10^{20}{\rm cm^{-2}}$) model atmosphere.  Curves show
the inferred $T$ for magnetic and non-magnetic H and Fe as a function
of the true $T_{\rm eff}$.}
\end{figure}

As in RR96, we can illustrate these broad band spectral effects by
simulating a modest statistic (3000 count) PSPC exposure of a neutron
star, whose absorbed spectrum is fitted by a blackbody. Comparing the
inferred $T_{BB}$ with the true $T_{\rm eff}$ for various atmosphere
models illustrates the low resolution spectral effects (Figure 5).  As
for non-magnetic atmospheres, the Fe results are much softer for a
given $T_{\rm eff}$ than H atmospheres. In contrast, existing data
suggest phase average spectra appreciably harder than the
corresponding blackbody. While the clear existence of magnetospheric
flux contributions and the likely $T_{\rm eff}$ variations over the
surface complicate the interpretation, it would seem that these data
support the existence of light element atmospheres on neutron star
surfaces. If substantiated, this conclusion would give important clues
to the history of the neutron star surface and serious implications
for cooling curves and the equation of state of matter at high
density. The magnetic model Fe atmospheres produced in this paper are
of sufficient quality to allow selection between H models and Fe
models for pulsar X-ray surface emission. If H atmospheres are
preferred, the sophisticated modeling possible for this simple species
will allow detailed interpretation of soft X-ray spectra. If however,
Fe models remain acceptable, further careful study of the behavior of
highly magnetized iron under neutron star surface conditions will be
strongly motivated.

\begin{acknowledgements} 
We thank G. G. Pavlov for supplying the sample magnetic H model spectra
and for giving a careful and detailed critique of this paper.
This work was supported in part by the National Sciences and
Engineering Research Council of Canada (MR); by NASA grants NAG
5-3101, and NAGW-4526 (RWR); and by NAG 5-2868 and, through the {\it
Compton} fellowship program, by NASA grant NAG 5-2687 (MCM).
\end{acknowledgements}

\end{document}